\documentclass[12pt]{article}

\usepackage{epsf}
\usepackage{latexsym,euscript}
\usepackage[dvips]{graphicx}
\usepackage{amsmath}
\usepackage{amsfonts}
\usepackage{amssymb, epsfig}
\usepackage{float}
\usepackage{array}

\usepackage{xcolor}
\usepackage{layout}

\usepackage{color,soul}

\DeclareMathOperator{\arcsinh}{arcsinh}

\begin{document}

\begin{center}
{\Large \textbf{Duals of lattice Abelian models with static 
determinant at finite density}}

\vspace*{0.6cm}
\textbf{O.~Borisenko${}^{\rm a}$\footnote{email: oleg@bitp.kiev.ua},
V.~Chelnokov${}^{\rm b}$\footnote{email: chelnokov@itp.uni-frankfurt.de, on leave from BITP},
S.~Voloshyn${}^{\rm a}$\footnote{email: billy.sunburn@gmail.com},
P.~Yefanov${}^{\rm c}$\footnote{email: paul.yefan@gmail.com}}

\vspace*{0.3cm}
{\large \textit{${}^{\rm a}$ N.N.Bogolyubov Institute for Theoretical
Physics, National Academy of Sciences of Ukraine, 03143 Kyiv, Ukraine}} \\ 
\vspace{0.3cm}
{\large \textit{${}^{\rm b}$ Institut f\"ur Theoretische Physik, Goethe-Universit\"at
Frankfurt, 60438 Frankfurt am Main, Germany}} \\ 
\vspace{0.3cm}
{\large \textit{${}^{\rm c}$ Department of Quantum Field Theory,
Taras Shevchenko National University of Kyiv, 03022 Kyiv, Ukraine}}
\end{center}

\begin{abstract}
Dual formulations of Abelian $U(1)$ and $Z(N)$ LGT with a static fermion determinant 
are constructed at finite temperatures and non-zero chemical potential. The dual form 
is valid for a broad class of lattice gauge actions, for arbitrary number of fermion 
flavors and in any dimension. The distinguished feature of the dual formulation is 
that the dual Boltzmann weight is strictly positive. This allows to gain reliable 
results at finite density via the Monte-Carlo simulations. As a byproduct of the dual 
representation we outline an exact solution for the partition function of the 
$(1+1)$-dimensional theory and reveal an existence of a phase with oscillating correlations. 
\end{abstract}

\section{Introduction}

There are many approaches designed to solve fully or partially the sign problem 
in QCD at finite chemical potential. One of such approaches is based on the dual 
representation for the partition function and physical observables.  The main idea 
is to perform an integration over original (gauge and fermion) degrees of freedom 
and to present the resulting weight in a positive form suitable for numerical 
simulations. A certain progress along this line of investigations has been achieved 
during last decade and can be briefly summarized as follows. 
The dual models with positive Boltzmann weights have been obtained and studied 
in Refs.\cite{Gattringer11,pl_dual20,Philipsen12,mcdual_21,3d_su3_lat21}. 
The calculations have been performed in the region of vanishing spatial gauge coupling 
constant and in the static approximation for the quark determinant (or at large quark masses).
In the strong coupling limit the $SU(N)$ LGT can be mapped onto 
monomer-dimer and closed baryon loop model \cite{karsch_89}. This dual representation 
has a soft sign problem and can be studied numerically. 
The positivity of the Boltzmann weight was also proven in the strong coupling 
limit of the scalar QCD with one, two or three scalar flavors \cite{scalar_qcd}.  
Beyond the strong coupling regime the dual formulation of $Z(3)$ gauge-Higgs model 
is also positive \cite{worm,z3lgt_boson} and suitable for Monte-Carlo simulations. 
Attempts to extend these results to full lattice QCD with the staggered fermions using 
different schemes of computations have not been so successful so far 
\cite{un_dual18,abelian-color-fluxes,dual-staggered-qcd}. 
Important result for the present paper was proven in Ref.\cite{2dqed}: 
the dual form of the massless two-dimensional $U(1)$ LGT with one or two flavors of staggered 
fermions is free of the sign problem and can be simulated with the help of a worm algorithm. 
Generalizing this result to a non-vanishing fermion mass proved to be a non-trivial task, and 
no solution has been found up to date. 

In this paper we extend results of 
Refs.\cite{Gattringer11,pl_dual20,Philipsen12,mcdual_21,3d_su3_lat21} on the dual formulation
to the case of arbitrary spatial gauge coupling for Abelian LGTs. The crucial simplification 
in dealing with Abelian models is the known exact and positive dual form 
of any $U(1)$ and $Z(N)$ pure gauge theory in any dimension. 
The purpose of this paper is to derive a positive dual formulation of Abelian LGTs 
with the full pure gauge action and arbitrary number of the staggered or Wilson flavors taken 
in the static approximation for the fermion determinants.  
As an application we discuss the possible updates of the dual Boltzmann weight appropriate 
for the Monte-Carlo simulations. Another direction we explore here is the solution of 
$(1+1)$-dimensional theory based on the dual representation. In particular, we calculate the 
eigenvalues of the corresponding transfer matrix and reveal the existence of an oscillating phase 
at finite density in all $Z(N)$ models with one or two fermion flavors.    

Our notations and conventions are as follows.  
We work on an anisotropic  periodic $(d+1)$-dimensional lattice $\Lambda = L^d\times N_t$ with 
spatial extension $L$ and temporal extension $N_t$. The lattice sites are denoted 
as $\vec{x}=(t,x)$ with $x=(x_1,\cdots,x_d)$, links in the temporal (spatial) direction are 
denoted as $l_t$ ($l_s$) and plaquettes as $p_t$ ($p_s$). The pure gauge action is of the form  
\begin{equation}
S_g(w_p) =  \beta_t \sum_{p_t} S(w_{p_t})  +  \beta_s \sum_{p_s} S(w_{p_s}) \ ,   
\label{gauge_action_def}
\end{equation}
where anisotropic coupling constants are related by $\beta_s = \beta_t \ \xi^2$  
with $\xi = \frac{a_t}{a_s}$. $a_t$ ($a_s$) is lattice spacing in the temporal (spatial) 
direction. $\beta = a_tN_t$ is an inverse temperature. 
The partition functions of $Z(N)$ and $U(1)$ LGTs are given by 
\begin{eqnarray} 
\label{zn_pf_def} 
Z_{\Lambda} &=& \sum_{\{ s_l \}=0}^{N-1} \ e^{S_g(s_p)} \ 
\prod_{f=1}^{N_f} \mbox{Det} {\cal{M}}_{\vec{x},\vec{x}^{\prime}} \ ,  \\ 
\label{u1_pf_def}
Z_{\Lambda} &=& \int_0^{2\pi} \prod_l \frac{d\phi_l}{2\pi}  \ e^{S_g(\phi_p)} \ 
\prod_{f=1}^{N_f} \mbox{Det} {\cal{M}}_{\vec{x},\vec{x}^{\prime}} \ , 
\end{eqnarray} 
where $s_p$ and $\phi_p$ are the standard plaquette angles.  
In the static approximation valid for large masses and/or for $\xi\ll 1$ 
the fermion determinant can be approximated as 
\begin{equation}
\mbox{Det} {\cal{M}}_{\vec{x},\vec{x}^{\prime}} \approx \prod_x \ 
A_f \ \left [ 1 + h_+^f W(x) \right ]^g \ \left [ 1 + h_-^f W^{\dagger}(x) \right ]^g \ . 
\label{stat_approx}
\end{equation} 
$g=1(2)$ for the staggered (Wilson) fermions, $W(x)=\prod_{t=1}^{N_t} U_0(t,x)$ 
is the Polyakov loop. The constants appearing on the right-hand side of (\ref{stat_approx}) 
are given by 
\begin{equation}
A_f = e^{2 N_t \arcsinh m_f} \ , \ 
h_{\pm}^f \ = \ e^{-(\arcsinh m_f \mp \mu_f) N_t} 
\label{hpm_stag}
\end{equation} 
for the staggered fermions and 
\begin{equation} 
A_f \ = \  (2\kappa_f)^{4 N_t} \ , \ 
h_{\pm}^f \ = \  \left ( 2\kappa_f \ e^{\pm \mu_f} \right )^{N_t} \ , 
\ \kappa_f \ = \ \frac{1}{2m_f+2 d +2\cosh\mu_f} 
\label{hpm_wilson}
\end{equation} 
for the Wilson fermions. In this paper we consider a class of ferromagnetic pure gauge actions
$S_g$ whose Boltzmann weight can be expanded as 
\begin{equation}
e^{S_g(\omega)} = \prod_p \ \sum_{r=-\infty}^{\infty} C_r \ e^{ir \omega} 
\label{u1_char_exp}  
\end{equation}
with positive coefficients $C_r$. {\it E.g.}, for the standard Wilson action one has 
$C_r=I_r(\beta)$, where $I_r(\beta)$ is the modified Bessel function.

\section{Dual representation} 

The Boltzmann weight of the models (\ref{zn_pf_def}) and (\ref{u1_pf_def}) 
is complex due to the fermion contribution (\ref{stat_approx}). 
It is straightforward to get a positive expression for this weight 
by integrating out explicitly all gauge degrees of freedom and rewriting the theory 
in terms of fermion and plaquette occupation numbers. 
In order to perform such integration, the static determinant with $N_f$ fermion flavors 
is presented as 
\begin{equation}
\sum_{\substack{k_1(x)=0\\k_1^{\prime}(x)=0}}^1 \ldots 
\sum_{\substack{k_{N_f}(x)=0\\k_{N_f}^{\prime}(x)=0}}^1 \ 
\prod_x \prod_{f=1}^{N_f} A_f \ (h_+^f)^{k_f(x)} \ (h_-^f)^{k_f^{\prime}(x)} \ 
\left ( W(x) \right )^{k_f(x)-k_f^{\prime}(x)} \ .
\label{u1_stat_det}
\end{equation}
Combining this representation with the expansion (\ref{u1_char_exp}) one can integrate 
over gauge fields to obtain, {\it e.g.} for the staggered fermions 
\begin{eqnarray}
Z &=& \sum_{\{r(p)\}=-\infty}^{\infty} \ \sum_{\substack{k_1(x)=0\\k_1^{\prime}(x)=0}}^1 \ldots 
\sum_{\substack{k_{N_f}(x)=0\\k_{N_f}^{\prime}(x)=0}}^1 \
\prod_{p_s} C_{r(p_s)}(\beta_s)  \prod_{p_t} C_{r(p_t)}(\beta_t) \nonumber \\ 
\label{pf_ab_Nf}
&\times&\prod_x \prod_{f=1}^{N_f} A_f \ (h_+^f)^{k_f(x)} \ (h_-^f)^{k_f^{\prime}(x)} \ 
\prod_{l_s} \delta_G \left ( \sum_{p\in l_s} \tilde{r}(p) \right ) \\ 
&\times&\prod_{l_t} \delta_G \left ( \sum_{p\in l_t} \tilde{r}(p) + 
\sum_{f=1}^{N_f} \left ( k_f(x)-k_f^{\prime}(x) \right ) \right ) \ . \nonumber 
\end{eqnarray}
In case of $N_f$ degenerate flavors the last expression simplifies to 
\begin{eqnarray}
&&Z = A^{g N_f L^d} \sum_{\{r(p)\}=-\infty}^{\infty} \ 
\sum_{\substack{k(x)=0\\k^{\prime}(x)=0}}^{g N_f} \
\prod_{p_s} C_{r(p_s)}(\beta_s)  \prod_{p_t} C_{r(p_t)}(\beta_t) 
\prod_{l_s} \delta_G \left ( \sum_{p\in l_s} \tilde{r}(p) \right )  \nonumber \\ 
&&\prod_{l_t} \delta_G \left ( \sum_{p\in l_t} \tilde{r}(p) + k(x)-k^{\prime}(x) \right ) \
\prod_x  \binom{g N_f}{k(x)} \binom{g N_f}{k^{\prime}(x)} \ 
h_+^{k(x)} \ h_-^{k^{\prime}(x)} \  .   
\label{pf_ab_Nf_deg}
\end{eqnarray} 
In the last equations $\delta_G(x)$ means the delta-function on the group $G=Z(N),U(1)$. 
Thus, the partition function is expressed in terms of fermion numbers $k_f(x),k_f^{\prime}(x)$ 
and plaquette occupation numbers $r(p)$. Both numbers are subject to constraints expressed via 
group delta-functions. The constraint on the spatial links $l_s$ is precisely the same as in 
the pure gauge theory due to the absence of spatial gauge fields in the fermion determinant. 
The constraint on the temporal links $l_t$ is modified due to a contribution of the Polyakov loops 
arising from the determinant. Note, fermion numbers $k_f(x),k_f^{\prime}(x)$ do not depend on the 
temporal coordinate $t$, {\it i.e.} they are equal for all time-like links with coordinates 
$l_t=(t,x;0)$ at fixed $x$. We have also used the following convention: $\tilde{r}(p)=r(p)$ if 
a given link $l_s$ or $l_t$ points in a positive direction when going around plaquette $p$ 
and $\tilde{r}(p)=-r(p)$, otherwise. 

As follows from the explicit representation of the group delta function $\delta_G(x)$ 
the dependence on $\mu$ drops out both from the partition function and from all invariant 
observables for $U(1)$ theory with one fermion flavor. To get a non-trivial dependence 
one has to consider a theory with $N_f\geq 2$ as in \cite{2dqed}. 
For $Z(N)$ model the dependence on chemical potential is non-trivial for any number of flavors.

It is straightforward to get dual representations for the most important observables. 
Taking into account Eq.(\ref{hpm_stag}) one obtains for the staggered fermions 
the particle density of $f$th flavor 
\begin{equation} 
B_f = \frac{1}{L^d N_t} \frac{\partial \ln Z}{\partial \mu_f} = 
\frac{1}{L^d} \left \langle \ \sum_x \left ( k_f(x)-k_f^{\prime}(x) \right ) \right \rangle \  
\label{part_density}
\end{equation} 
and the fermion condensate of $f$th flavor 
\begin{equation} 
\sigma_f = \frac{1}{L^d N_t} \frac{\partial \ln Z}{\partial m_f} = 
\frac{1}{\sqrt{1+m_f^2} L^d} \left \langle 
\sum_x \left ( 2- k_f(x) - k_f^{\prime}(x) \right ) \right \rangle \ . 
\label{ferm_condensate}
\end{equation} 
Extension to the Wilson fermions is trivial. Plaquette expectation value is 
\begin{equation} 
P(p) = \frac{1}{2} \ \left \langle  
\frac{I_{r(p)-1}(b)+I_{r(p)+1}(b)}{I_{r(p)}(b)}  \right \rangle \ , 
\label{plaq_dual}
\end{equation} 
where $b=\beta_s (\beta_t)$ stays for the spatial (temporal) plaquette. 
Expectation value of the pure gauge action becomes 
\begin{equation} 
\left \langle S_g \right \rangle = \frac{1}{L^d N_t} \ \left ( 
\beta_s \sum_{p_s} P(p_s) + \beta_t \sum_{p_t} P(p_t) \right ) \ . 
\label{gauge_action__dual}
\end{equation} 
Correlation functions of the Polyakov loops can be calculated as a ratio of the partition 
functions 
\begin{equation} 
\left \langle W(x) W^{*}(y) \right \rangle = \frac{Z(\eta_x,\bar{\eta}_y)}{Z} \ .   
\label{PL_corr_dual}
\end{equation} 
The partition function $Z(\eta_x,\bar{\eta}_y)$ coincides with $Z$ up to a modification 
of the delta's on all temporal links with coordinates $l=(t,x;0)$ and $l=(t,y;0)$: arguments 
of these delta-functions acquire a linear shift by $\eta=-\bar{\eta}=1$.  

The models defined in Eqs.(\ref{pf_ab_Nf}) and (\ref{pf_ab_Nf_deg}) have explicitly 
non-negative weights (for $h_+, h_- > 0$), hence they can in principle be studied  with 
numerical Monte-Carlo simulations.  
The delta functions in the partition function create constraints on the configurations, 
which have to be preserved by the updates. As a first approach to the numerical simulation 
we propose a Metropolis algorithm for a $(d+1)$ model on a lattice with periodic 
boundary conditions, that attempts following updates:
\begin{itemize}
    \item For $Z(N)$ models - change of each variable 
    ($k_f(x)$, $k_f^\prime(x)$, $r(p)$) by $\pm N$.
    
    \item Change of two $k$ variables at the same $x$ by $\pm 1$, preserving the sum 
          $\sum_f (k_f(x) - k_f^\prime(x))$.
          
    \item Change by $\pm 1$ of two $k$ variables at two neighboring space positions 
          $x$, $y$, compensated by the corresponding change at each time-like 
          plaquette between sites $x$ and $y$.
          
    \item Change by $\pm 1$ of $r(p)$ variables on plaquettes forming a unit three dimensional
          cube.
          
    \item Global change by $\pm 1$ of all $r(p)$ variables in $\mu \nu$ direction forming 
    a surface wrapping around the whole lattice.
\end{itemize}
These updates generate the full set of permitted configurations, though it is possible that two 
configurations with large weight are connected through configurations with much smaller weight, 
which would reduce the update algorithm efficiency. 
A more efficient alternative would be to develop a surface-building worm update algorithm 
similar to the ones proposed in \cite{worm}.  

Another approach is to get rid of constraints on configurations whenever possible. 
First, consider the representation (\ref{pf_ab_Nf_deg}) for $(2+1)$-dimensional $U(1)$ theory. 
When $k(x) = k^\prime(x) = 0$ we recover the dual representation for the pure gauge model.  
The solution of the constraint is well known and reads \cite{dualu1} 
\begin{equation}
\label{d2_gauge_sol}
\tilde{r}(p) = q(x) - q(x+e_{\nu})  \ . 
\end{equation}
$q(x)$ is a new set of integer variables defined in the sites of the dual lattice. 
We have neglected some global variables. These global variables are conjugate to 
global Bianchi identities and do not contribute to thermodynamic limit. 
Possibility of nonzero $k$ is restored by modifying the conditions for the temporal plaquettes:
\begin{align}
    \label{d2_sol_ps}
    \tilde{r}(p_s) & = q(x) - q(x+e_0) \ , \\
    \label{d2_sol_pt}
    \tilde{r}(p_t) & = q(x) - q(x+e_n) + \rho_n(x) \ , \ n=1,2 \ .  
\end{align}
Here $\rho_n(x)$ are new integer variables defined on the dual links $(x,n)$ and depending only 
on spatial coordinates. 
Substituting Eqs.~(\ref{d2_sol_ps}), (\ref{d2_sol_pt}) into our constraints we see that 
the constraints on $l_s$ are satisfied, while the constraints on $l_t$ appear only 
at one fixed time slice and read
\begin{equation}
\label{2d_delta_x}
\delta_G \left ( \rho(p) - k(p) + k^{\prime}(p) \right ) \ , \ 
\rho(p) = \rho_1(x) + \rho_2(x + e_1) - \rho_1(x+e_2) - \rho_2(x) \ . 
\end{equation}
Four links entering this constraint form a dual plaquette $p$.
Since now each of the variables $k(p)$, $k^\prime(p)$ 
appear just in one delta function, and the terms in partition function that depend on $k$ 
do not mix at different plaquettes, we can calculate the sum over
$k(p)$ and $k^\prime(p)$ at each plaquette to remove the last set of deltas.  
This leads to the following dual form of the partition function 
\begin{align}
\label{pf_u1_2d_solved}
Z &= A^{g N_f L^d} \sum_{\{q(x)\}=-\infty}^{\infty} \  
\prod_{l_t} C_{q(x) - q(x+e_0)}(\beta_s)  
\nonumber \\
& \times 
\sum_{\rho_n(x)=-\infty}^{\infty} \ \prod_{l_s} C_{q(x) - q(x+e_n) + \rho_n(x)}(\beta_t) 
\prod_p K_{\rho(p)} \ , \\
\label{pf_u1_2d_K}
K_\rho & =\left(\frac{h_+}{h_-}\right)^\frac{\rho}{2}\frac{(gN_f)!}{(gN_f+\rho)!} 
P^\rho_{gN_f}\left(\frac{1+h_+h_-}{1-h_+h_-}\right) \  , 
\end{align} 
where $P^{\rho}_n(x)$ is the associated Legendre function. 
Product $\prod_p$ runs over all space-like plaquettes of the dual lattice at a fixed time slice. 
To simulate the model (\ref{pf_u1_2d_solved}) one can precompute 
$K_\rho$ for $-g N_f \leq \rho \leq g N_f$ and then perform Metropolis updates 
by $\pm 1$ on each variable $q(x)$, $\rho_n(x)$. 

For $N_f$ non-degenerate flavors, Eq.(\ref{pf_ab_Nf}), the representation (\ref{pf_u1_2d_solved}) 
remains valid. The only change is the expression for $K_\rho$ which becomes more complicated. 
An extension to $d=3$ theory can be accomplished in a similar way if one uses the solution 
of the constraint for the pure gauge model following \cite{dualu1}.
Finally, the $Z(N)$ case is recovered by treating each solution as an equality modulo $N$, 
thus leaving a degree of freedom for the difference of left and right parts divided by $N$.

\section{$(1+1)$-dimensional theory}

As an application, let us consider the dual formulation in $(1+1)$-dimensions. 
Due to deltas on spatial links $l_s$ all plaquette numbers at fixed position $x$ are equal 
and can be identified with a link variable $r(l)$ of a one-dimensional lattice. 
All deltas on temporal links with a fixed coordinate $x$ become also equal and can be associated 
with a site $x$ of the same one-dimensional lattice. The $U(1)$ partition 
function (\ref{pf_ab_Nf}) gets the form ($\beta_s=\beta_t=\beta$)
\begin{eqnarray}
Z &=& \sum_{\{r(l)\}=-\infty}^{\infty} \ \sum_{\substack{k_1(x)=0\\k_1^{\prime}(x)=0}}^1 \ldots 
\sum_{\substack{k_{N_f}(x)=0\\k_{N_f}^{\prime}(x)=0}}^1 \ 
\prod_{l} C_{r(l)}^{N_t}(\beta) \ 
\prod_x \prod_{f=1}^{N_f} A_f \ (h_+^f)^{k_f(x)} \ (h_-^f)^{k_f^{\prime}(x)}  \nonumber \\ 
\label{pf_2du1_def}
&\times&\prod_{x} \delta_G \left (r(l)-r(l-1) + 
\sum_{f=1}^{N_f} \left ( k_f(x)-k_f^{\prime}(x) \right ) \right ) \ . 
\end{eqnarray}
For $Z(N)$ model one has to make the following replacement in the last expression 
\begin{equation} 
\sum_{r=-\infty}^{\infty} \rightarrow \sum_{r=0}^N \sum_{q=-\infty}^{\infty} \ , \ \ 
 C_r(\beta) \rightarrow C_{r+q N}(\beta) \ .
\label{pf_2dzn_def}
\end{equation} 
This partition function can be evaluated as 
\begin{equation}
Z = C_0^{L N_t}(\beta) \ \prod_{f=1}^{N_f} A_f^L \ \sum_{i=0} \ \lambda_i^L \ , 
\label{pf_2d_solution}
\end{equation} 
where $\lambda_i$ are eigenvalues of the following transfer matrix 
\begin{eqnarray}
T_{r_1 r_2} = \sqrt{B_{r_1} B_{r_2}} \ \sum_{\substack{k_1=0\\k_1^{\prime}=0}}^1 \ldots
\sum_{\substack{k_{N_f}=0\\k_{N_f}^{\prime}=0}}^1 \ 
\prod_{f=1}^{N_f} (h_+^f)^{k_f} \ (h_-^f)^{k_f^{\prime}} \ , 
\label{transf_matr_2d}
\end{eqnarray} 
where $B_{r}=C_r^{N_t}(\beta)/C_0^{N_t}(\beta)$ and all configurations are subject to constraint 
$r_1-r_2 + \sum_{f=1}^{N_f} \left ( k_f-k_f^{\prime} \right ) = 0 (\mbox{mod} N)$. 
Below we analyze the theory with the Wilson action and two staggered fermion flavors. 

\begin{figure}[ht]
\centerline{{\epsfxsize=6.5cm \epsfbox{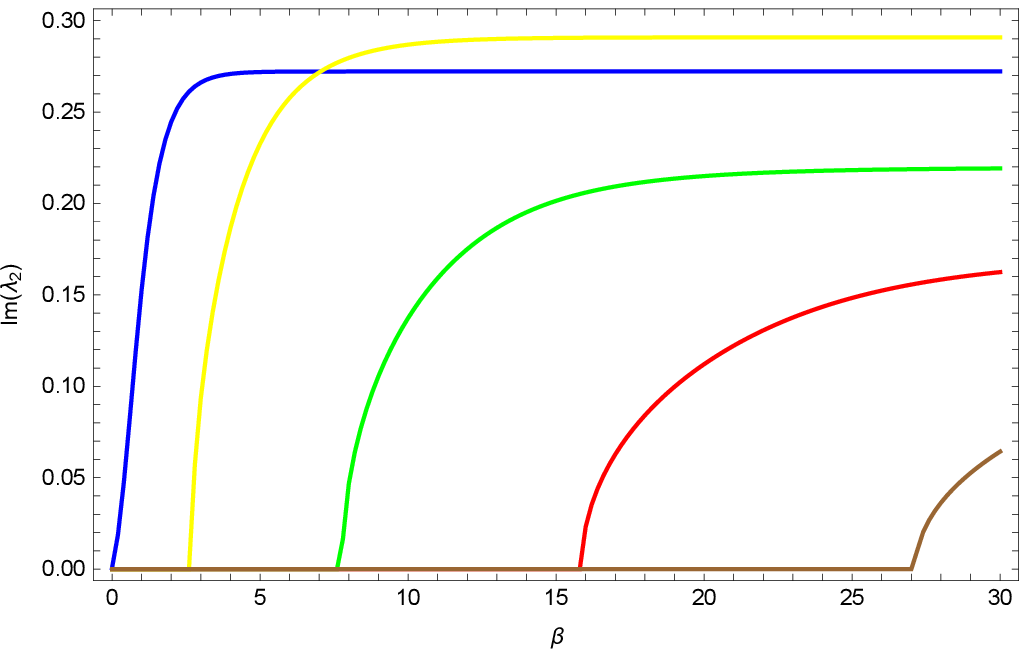}} {\epsfxsize=6.5cm\epsfbox{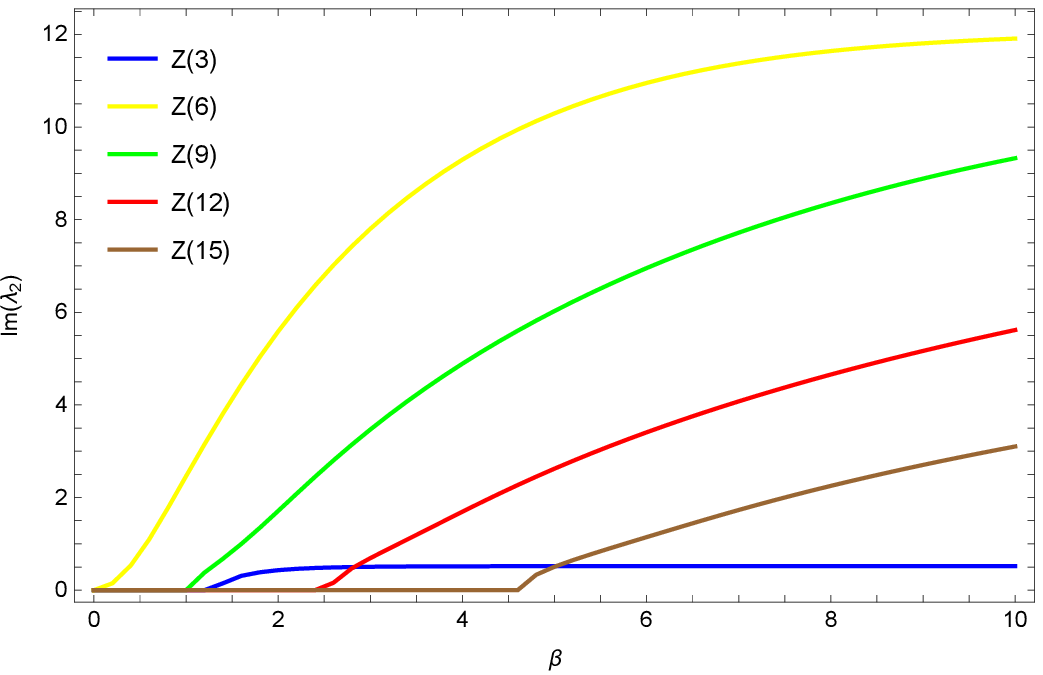}}}
\caption{\label{Im_lambda} Plots of the imaginary part of the 2nd eigenvalue of the transfer 
matrix of $Z(N)$ model with two flavors of staggered fermions as a function of the coupling 
constant. Left panel: $m_1 = 3, m_2 = 1$, $\mu_1 = -0.16, \mu_2 = 0.45$. 
Right panel: $m_1 = 1, m_2 = 0.1$, $\mu_1 = 2, \mu_2 = 1$. }
\end{figure} 

When chemical potentials are zero all eigenvalues are real. This leads to a familiar exponential 
decay of the connected part of the Polyakov loop correlation function. However, when non-zero 
chemical potentials are introduced, one finds such values of the coupling constant above which 
the eigenvalues become complex. Moreover, the second and the third eigenvalues are conjugate 
to each other. Typical examples of such behavior are shown in Fig.\ref{Im_lambda} for various 
values of $N$. This implies the following decay of the two-point correlation function 
\begin{equation}
\langle W(0) W^*(R) \rangle_c \approx e^{-m_r R} \cos m_i R \ .   
\label{2point_decay}
\end{equation}  
Such an oscillating decay should not come as a surprise. Indeed, in a similar settings 
it was found in the $(1+1)$-dimensional $SU(3)$ theory with one flavor \cite{ogilvie2016} 
and in the two-dimensional $Z(3)$ spin model in a complex magnetic field \cite{forcrand_z3} 
as well as in the 't Hooft-Veneziano limit of $SU(N)$ Polyakov loop models \cite{largeN_2021}.  
In all cases studied we have found the increase of the $\beta$ value with $N$ above which 
the oscillating phase appears. We do not know if the values of masses and/or chemical potentials 
can be re-scaled in a way such that in the limit $N\to\infty$ the oscillating phase would exist. 
We have, however studied $U(1)$ model directly in the region $\beta\leq 10$ and various values 
of masses and chemical potentials. No oscillating phase have been found in this case. 
We thus think the reasonable conjecture is to assume that the complex spectrum of the eigenvalues 
does not appear in the $U(1)$ model with two fermion flavors though this issue requires more 
thorough investigation.

\section{Summary} 

In this paper we have derived the dual representations for $U(1)$ and $Z(N)$ lattice gauge theories  
in $(d+1)$-dimension with $N_f$ staggered or Wilson fermion flavors and in the static 
approximation for the fermion determinant. We presented two essentially different representations: 
one with a set of constraints on the dual variables, the second one is free of constraints. 
In both cases the dual weight is positive and suitable for numerical simulations. Even the dual 
model with constraints can be studied numerically if the proper algorithm is developed. 
One such possible algorithm was suggested here. 
As an application of the dual form we have studied $(1+1)$-dimensional model with two staggered 
flavors. The model can be solved with the help of the transfer matrix. This solution reveals 
an existence of a phase in all $Z(N)$ models with an exponential decay of the correlations 
modulated by an oscillating function. The value of the coupling constant, above which such 
phase appears, grows with $N$.   

Further possible applications of the dual formulation would be to study 1) the large 
$N_f$ limit of Abelian models at finite density and 2) the Berezinskii-Kosterlitz-Thouless 
phase transition in $2+1$ models. The dual formulation of $U(1)$ model turned out to be very 
efficient in the study of this type of phase transition in a pure gauge model \cite{bkt_3du1}. 
We think it can be also useful to investigate how the finite-density affects the critical 
behavior. These problems are currently under investigation.

Probably, the most important question is whether this approach can be extended to the full 
fermion determinant. On our opinion, combining the present approach with the methods of 
Ref.\cite{2dqed} one could construct the positive dual weight for $(1+1)$ dimensional Abelian 
models with non-zero fermion masses. This possibility certainly deserves further investigations.

\vspace*{0.5cm}

O. Borisenko acknowledges support from the National 
Academy of Sciences of Ukraine in frames of the project 
"Meeting new experimental data on proton-proton, proton-nuclei, 
nuclei-nuclei interactions at high energies at CERN, BNL, 
FERMILAB, GSI for theoretical analysis" (No. 0121U112254). 
V. Chelnokov acknowledges support by the Deutsche
Forschungsgemeinschaft (DFG, German Research Foundation) through the CRC-TR 211 
"Strong-interaction matter under extreme conditions" – project number
315477589 – TRR 211.


\begin{thebibliography}{99}

%
\bibitem{Gattringer11} C.~Gattringer, Nucl.Phys. B \textbf{850} (2011) 242 
[arXiv:1104.2503 [hep-lat]].
%
\bibitem{pl_dual20} O.~Borisenko, V.~Chelnokov, S.~Voloshyn, Phys.Rev. D {\bf 102} (2020) 
014502 [arXiv:2005.11073 [hep-lat]]. 
%
\bibitem{Philipsen12} M.~Fromm, J.~Langelage, S.~Lottini, O.~Philipsen,
JHEP \textbf{01} 042 (2012) [arXiv:1111.4953 [hep-lat]].
%
\bibitem{mcdual_21} 
O. Borisenko, V. Chelnokov, E. Mendicelli, A. Papa, Nucl.Phys.B {\bf 965} (2021) 115332  
[arXiv:2011.08285 [hep-lat]].
%
\bibitem{3d_su3_lat21} O.~Borisenko, V.~Chelnokov, E.~Mendicelli, A.~Papa, 
Proceedings of Science, PoS(LATTICE2021) 587, [arXiv:2112.00043 [hep-lat]].
%
\bibitem{karsch_89} F.~Karsch, K.H.~M\"{u}tter, Nucl.Phys. B {\bf 313} (1989) 541. 
%
\bibitem{scalar_qcd} F.~Bruckmann, J.~Wellnhofer,  Phys.Rev. D {\bf 97} (2018) 014501 
[arXiv:1710.08243 [hep-lat]]. 
%
\bibitem{worm} Y.~Delgado, C.~Gattringer, A.~Schmidt, Comput. Phys. Commun. {\bf 184} (2013) 1535, 
[arXiv:1211.3436 [hep-lat]].
%
\bibitem{z3lgt_boson} K.~Langfeld, String-like theory as solution to the sign problem 
of a finite density gauge theory, PoS (Confinement2018) 049, [arXiv:1811.12921 [hep-lat]]. 
%
\bibitem{un_dual18} O.~Borisenko, V.~Chelnokov, S.~Voloshyn, EPJ Web Conf. {\bf 175} (2018) 
11021 [arXiv:1712.03064 [hep-lat]].
%
\bibitem{abelian-color-fluxes} C.~Marchis, C.~Gattringer, Phys.Rev. D {\bf 97} (2018) 034508 
[arXiv:1712.07546 [hep-lat]]. 
%
\bibitem{dual-staggered-qcd} G.~Gagliardi, W.~Unger, Phys.Rev. D {\bf 101} (2020) 034509 
[arXiv:1911.08389 [hep-lat]]. 
%
\bibitem{2dqed} C.~Gattringer, T.~Kloiber, V.~Sazonov, Nucl.Phys. B {\bf 897} (2015) 732 
[arXiv:1502.05479 [hep-lat]]. 
%
\bibitem{dualu1} T.~Banks, J.~Kogut, R.~Myerson, Nucl.Phys. B {\bf 121} (1977) 493.
%
\bibitem{ogilvie2016} H.~Nishimura, M.~Ogilvie, K.~Pangeni, 
Phys.Rev. D \textbf{93} (2016) 094501 [arXiv:1512.09131 [hep-lat]]. 
%
\bibitem{forcrand_z3} O.~Akerlund, P.~de Forcrand, T.~Rindlisbacher, 
JHEP \textbf{10} (2016) 055 [arXiv:1602.02925 [hep-lat]]. 
%
\bibitem{largeN_2021} O.~Borisenko, V.~Chelnokov, S.~Voloshin,  Proceedings of Science, 
PoS(LATTICE2021) 453, [arXiv:2111.07103 [hep-lat]].
%
\bibitem{bkt_3du1} O.~Borisenko, V.~Chelnokov, M.~Gravina, A.~Papa, JHEP {\bf 09} 
(2015) 062 [arXiv:1507.00833 [hep-lat]]. 


\end{thebibliography}
\end{document}